# Rapid sampling of all-atom peptides using a library-based polymer-growth approach

*Artem B. Mamonov[1], Xin Zhang[2], and Daniel M. Zuckerman[1]*

[1] Department of Computational Biology, School of Medicine, University of Pittsburgh, Pittsburgh, Pennsylvania 15260

[2] Department of Physics and Astronomy, University of Pittsburgh, Pittsburgh, Pennsylvania 15260

**Abstract**

We adapted existing polymer growth strategies for equilibrium sampling of peptides described by modern atomistic forcefields with implicit solvent. The main novel feature of our approach is the use of pre-calculated statistical libraries of molecular fragments. A molecule is sampled by combining fragment configurations – of single residues in this study – which are stored in the libraries. Ensembles generated from the independent libraries are reweighted to conform with the Boltzmann factor distribution of the forcefield describing the full molecule. In this way, high-quality equilibrium sampling of small peptides (4-8 residues) typically requires less than one hour of single-processor wallclock time and can be significantly faster than Langevin simulations. Furthermore, approximate but clash-free ensembles can be generated for larger peptides (e.g., 16 residues) in less than a minute of single-processor computing. We also describe an application to free energy calculation, a "multi-resolution" implementation of the growth procedure and application to fragment assembly protein-structure prediction protocols.

## I. INTRODUCTION

This paper investigates whether decades-old polymer-growth algorithms [1-14] have promise for the study of biomolecules modeled by modern atomistic forcefields. Although polymer approaches have previously been applied to peptides [15-17], their application to atomistic forcefields at physiological temperatures has been problematic [18-20]. Here we report a novel implementation of growth algorithm based on pre-calculated statistical libraries of molecular fragment configurations and energies. The encouraging results from a limited set of small test peptides, reported below, suggest that further investigation is warranted.

The well-known problem of sampling biomolecules typically has been addressed by dynamical simulations and variants – molecular dynamics (MD), Langevin dynamics (LD), and Metropolis Monte Carlo with local moves. All these approaches suffer from the well-known problem of undersampling: dynamical simulations of proteins are far too short to probe timescales (and motions) thought to be of dominant biological importance. Even simulations of modest-sized peptides are slow to "converge" [21,22]. Sophisticated variants of dynamical simulations, such as replica exchange [23-26], also have not convincingly solved the undersampling problem [27-29]. While multi-resolution methods appear to have substantial promise [30-33], rigorous applications have been restricted to small systems thus far.

The importance of sampling biomolecules and the intrinsic limitations of dynamical simulation together suggest the value of exploring fully non-dynamical polymer growth algorithms. Such methods have a history dating back more than fifty years. Initial studies focused on straightforward build-up of lattice-polymer chains [1,3,4], but the early approaches were limited by the "attrition problem," in which the vast majority of chains encounter dead ends before reaching a significant size. Our own approach builds directly on methods developed to treat attrition, especially (i) the Rosenbluths approach of re-weighting chains based on possible growth steps [5], and (ii) equally seminal work by Wall and Erpenbeck describing "enrichment" of successful partially grown chains by replication and appropriate weighting [7]. Wall, Rubin and Isaacson noted that future increments of the growth of a lattice polymer were limited to a small set of configurations [6], partly anticipating the libraries we employ here. Many additional improvements have also been proposed [8-10]. The basic theory behind polymer growth as we apply it, along with key practical insights, was fully set out by Garel and Orland in 1990 [11]. Important descriptions of growth algorithms are also provided by Grassberger [12,13] and by Liu [14].

Polymer growth algorithms have been applied previously to biomolecules. Highly simplified models of proteins were studied by Grassberger and coworkers [15] and by Liu and coworkers [16,17,34]. Garel, Orland, and coworkers applied polymer growth methods to all-atom peptide models — but their work employed extremely high-temperature sampling (T=1000 K) followed by energy minimization [11,18-20]. Our use of pre-calculated fragment libraries emulates ideas from the ROSETTA software [35] as well as from work by Clementi and coworkers [36,37]. However, none of these previous studies appears to have generated canonical sampling for a modern atomistic forcefield at T ~ 300 K.

In light of the significant body of historical work, the present contribution must be considered pragmatic rather than theoretical. In brief, our work shows that pre-generated libraries of statistically distributed monomer/fragment configurations can be used in implicit solvent sampling of all-atom molecular systems at temperatures of interest (T = 300 K). For high quality statistical sampling the present implementation is limited to small peptides – up to about eight residues and less than 100 atoms. However, besides equilibrium sampling, our growth procedure can be also used for rapid generation of approximate (i.e., steric-clash free) ensembles of larger peptides containing up to ca. 16 amino acid residues. Although the present work is formally similar to our previous use of fragments for free energy calculations [38], this study presents critical technique improvements which greatly improve efficiency.

Our study also employs recently developed statistical approaches [39] to quantify the degree to which efficiency has been gained. The library-based strategy is shown to be extremely efficient in some cases — decreasing the required wallclock time by over one order of magnitude. However, we believe that several improvements are possible, as described in the Discussion section.

In our approach the choice of fragments is flexible and they can correspond to different groups of atoms in the molecule. For proteins the natural choice of fragments is the amino acid residues because proteins consist of only 20 building blocks. However, other choices are possible. When the fragments correspond to the backbone and side chains, the procedure is essentially a multi-resolution method. The backbone can be sampled using other methods such as our previously developed library-based Monte

Carlo [40], followed by the gradual addition of more atomistic detail embodied in side chains.

## II. FORMALISM

As noted in the Introduction, polymer growth algorithms have been developed and used over decades [1-14]. Our approach follows earlier work in many regards, but is specifically tailored to the use of modern atomistic forcefields and implicit solvent. Our presentation of the algorithms relies solely on straightforward re-weighting concepts [14,41]. We describe a simple and apparently novel approach to using libraries of molecular fragments which can save significant computational cost.

### II.A. Forcefield, fragments and notation

In this study we generate equilibrium configurations according to the OPLS-AA forcefield [42] using a simple implicit solvent model (with uniform dielectric constant of 60) at 298 K. This dielectric constant has been chosen to give reasonable agreement for Ramachandran propensities as compared to GBSA solvent model [43].

The potential energy of the forcefield plus the solvent model will be denoted by $U(\mathbf{x})$, where the full set of 3$N$-6 internal coordinates $\mathbf{x} = (x_1, x_2, ..., x_{3N-6})$, consists of $N$-1 bond, $N$-2 bond angles and $N$-3 dihedrals. The full set of coordinates corresponding to a single molecular fragment $y$ will be denoted by $\mathbf{x}_y$ with $y = A, B, C, ...$. The collection of forcefield terms for fragment $y$, denoted by $U_y$ will contain all terms internal to the particular subset of atoms included in the fragment. That is, it will include all bonded and non-bonded terms for those atoms. Dummy atoms may be added to a fragment, as in the present study, to include the six degrees of freedom that specify the orientation of fragments relative to each other. However, dummy atoms will have no effect on the trial distribution.

We assume that fragments are non-overlapping and exactly divide all coordinates, so that for the whole molecule the full set of coordinates may be written as

$$\mathbf{x} = \mathbf{x}_A, \mathbf{x}_B, \mathbf{x}_C, \ldots \qquad (1)$$

It is important to realize that the full forcefield $U$ can never be written as a sum of fragment forcefields $U_y(\mathbf{x}_y)$. The reason is that, regardless of which intermediate coordinates are included via dummy atoms, no coordinate set $\mathbf{x}_y$ includes distances between atoms from different fragments. Needless to say, such inter-atomic distances are fundamental to the full molecular forcefield. Inter-fragment interactions are fully accounted for in our growth procedure, as described below.

**II.B. Combination of fragments**

In our approach, a molecule is sampled by growing it from scratch using pre-calculated molecular fragments. Here we describe the process of joining fragments which may be repeated inductively by adding additional "monomers" onto the growing chain. Configurations for each fragment are calculated in advance so that they are distributed according to the Boltzmann factor of the forcefield describing the fragment. The set of Boltzmann-distributed configurations for each fragment is called a "library".

The basic procedure for joining fragments is simple. A new fragment configuration is drawn with uniform probability from its library and added to the partially grown chain. The interaction energy between the new fragment and other previously added fragments is evaluated. The generated configurations are reweighted to the Boltzmann factor distribution describing the partially grown molecule to correct for the new interactions.

Consistent with free energy calculations using our growth process [38], we will define a set of intermediate models $U_j$ which correspond to different stages of the growth process. We note that these intermediates are a little different than employed (before) in ref [38].

For a molecule consisting of *k* fragments, we will employ *k* intermediate models with interactions between fragments gradually "turned on". The first intermediate $U_1$, sampled at the library generation stage, includes interactions internal to each fragment, while subsequent intermediates add the indicated interactions among fragments $A, B, C, \ldots$. These intermediate models can be written as

$$\begin{aligned}
U_1(\mathbf{x}) &= U_A(\mathbf{x}_A) + U_B(\mathbf{x}_A) + U_C(\mathbf{x}_C) + \ldots \\
U_2(\mathbf{x}) &= U_1(\mathbf{x}) + U_{AB}(\mathbf{x}_A, \mathbf{x}_B) \\
U_3(\mathbf{x}) &= U_2(\mathbf{x}) + U_{AC}(\mathbf{x}_A, \mathbf{x}_C) + U_{BC}(\mathbf{x}_B, \mathbf{x}_C), \\
&\ldots \\
U(\mathbf{x}) &= U_{k-1}(\mathbf{x}) + \sum_{y=A,B,\ldots} U_{yz}(\mathbf{x}_{yz})
\end{aligned} \qquad (2)$$

where $U_{yz}$ denotes all forcefield interaction terms between fragments *y* and *z*. The last intermediate $U(\mathbf{x})$ is simply the full molecule and the sum $\sum_{y=A,B,\ldots} U_{yz}(\mathbf{x}_{yz})$ represent interactions between the last fragment *z* and all other fragments in the molecule.

**II.C. Growth by reweighting**

Our polymer-growth approach heavily relies on the re-weighting concept [14,41] because interactions between fragments are not included in the libraries of individual fragments. In essence we generate configurations with non-interacting fragments and gradually reweight them into an ensemble with all interactions. In other words the purpose of reweighting is to effectively put back all the interactions and correlations between fragments into the molecule.

At each stage, we want to generate a suitably distributed ensemble – called the target ensemble $P_j^{\text{targ}} \propto \exp\left[-\beta U_j(\mathbf{x})\right]$ for stage $j$ with the set $U_j$ defined in Eq. (2). When $j<k$, this target ensemble based on $U_j$ includes interactions only for the partially "grown" molecule. Yet configurations for stage $j$, as will be seen, are generated according to a different distribution, denoted $P_j^{\text{gen}}$. Hence, configurations must be *reweighted* according to

$$u_j(\mathbf{x}) = \frac{P_j^{\text{targ}}(\mathbf{x})}{P_j^{\text{gen}}(\mathbf{x})}, \tag{3}$$

where $u_j(\mathbf{x})$ is the weight of a configuration at stage $j$. (In fact, as explained below, $u_j(\mathbf{x})$ is an intermediate weight.) In Eq. (3) and subsequent equations, the symbol $\mathbf{x}$ does indeed represent the full set of coordinates. In intermediate stages $j<k$, however, some interactions are omitted: see Eq. (2).

To perform the reweighing procedure, we need to define the $P^{\text{gen}}$ and $P^{\text{targ}}$ for each intermediate stage. Let us consider each stage in detail. The first stage $U_1$ includes interactions within each fragment which are sampled at the library generation stage. The second stage $U_2$ corresponds to turning on interactions between fragments $A$ and $B$, starting from configurations already distributed according to $U_1$. Thus the generating distribution $P_2^{\text{gen}}$ is simply proportional to the Boltzmann factor describing the first intermediate with non-interacting fragments:

$$P_2^{\text{gen}}(\mathbf{x}) \propto \exp{-\beta U_1(\mathbf{x})}. \tag{4}$$

The distribution targeted at the second stage $P_2^{\text{targ}}$ proportional to the Boltzmann factor describing the second intermediate:

$$P_2^{\text{targ}}(\mathbf{x}) \propto \exp{-\beta U_2(\mathbf{x})}. \tag{5}$$

At the third stage, similarly, interactions are turned on between fragment $C$ and previously combined fragments $A$ and $B$. As before $P_3^{\text{gen}}$ is nothing but $P_2^{\text{targ}}$

$$P_3^{\text{gen}}(\mathbf{x}) = P_2^{\text{targ}}(\mathbf{x}) \propto \exp{-\beta U_2(\mathbf{x})}. \tag{6}$$

Likewise, $P_3^{\text{targ}}$ distribution is proportional to the Boltzmann factor describing the third intermediate:

$$P_3^{\text{targ}}(\mathbf{x}) \propto \exp{-\beta U_3(\mathbf{x})}. \tag{7}$$

It is not difficult to generalize this combination process for any other intermediate. For the *k*th intermediate (corresponding to the full molecule) $P_j^{\text{gen}}$ and $P_j^{\text{targ}}$ can be written as

$$P_k^{\text{gen}}(\mathbf{x}) = P_{k-1}^{\text{targ}}(\mathbf{x}) \propto \exp{-\beta U_{k-1}(\mathbf{x})} \tag{8}$$

$$P_k^{\text{targ}}(\mathbf{x}) \propto \exp{-\beta U(\mathbf{x})} \ . \tag{9}$$

It is important to note that in our procedure $P^{\text{gen}}$ is built sequentially based on $P^{\text{targ}}$ from the previous stages. This is the essence of "sequential importance sampling" [14] i.e., the probability distribution of the full molecule is built sequentially step by step. The advantage of sequential importance sampling is that the probability distribution is changed in small increments to give the better overlap between $P^{\text{gen}}$ and $P^{\text{targ}}$ at each stage.

The required partial weights $w_j$ can be calculated based on the incremental weights of Eq. (3). Specifically, the weight of a configuration at stage $j$ can be written recursively based on the weights from previous stages:

$$w_j = w_{j-1} u_j. \tag{10}$$

Substituting the corresponding $P^{\text{gen}}$ and $P^{\text{targ}}$ from Eqs. (4)-(9) into Eq. (10) the partial weights can be written as

$$\begin{aligned}
w_1(\mathbf{x}) &= 1 \\
w_2(\mathbf{x}) &\propto w_1(\mathbf{x}) \frac{\exp{-\beta U_2(\mathbf{x})}}{\exp{-\beta U_1(\mathbf{x})}} = w_1(\mathbf{x}) \exp{-\beta U_{AB}(\mathbf{x}_A, \mathbf{x}_B)} \\
w_3(\mathbf{x}) &\propto w_2(\mathbf{x}) \frac{\exp{-\beta U_3(\mathbf{x})}}{\exp{-\beta U_2(\mathbf{x})}} = w_2(\mathbf{x}) \exp\left[-\beta \left(U_{AC}(\mathbf{x}_A, \mathbf{x}_C) + U_{BC}(\mathbf{x}_B, \mathbf{x}_C)\right)\right], \\
&\ldots \\
w(\mathbf{x}) &\propto w_{k-1}(\mathbf{x}) \frac{\exp{-\beta U(\mathbf{x})}}{\exp{-\beta U_{k-1}(\mathbf{x}_{k-1})}} = w_{k-1}(\mathbf{x}) \exp\left[-\beta \sum_{y=A,B,\ldots} U_{yz}(\mathbf{x}_y, \mathbf{x}_z)\right]
\end{aligned} \tag{11}$$

where $w(\mathbf{x})$ is the total weight for the full molecule i.e., with interactions "turned on" between all fragments. Note that $w_1(\mathbf{x})$ is equal to one by construction because fragment configurations in the libraries are distributed according to the corresponding $P^{\text{targ}}$ – i.e., the Boltzmann factor describing the individual fragments.

Our "resampling" protocol, described later, will use the partial weights $w_j$. However, it is instructive to note that the total weight $w(\mathbf{x})$ in Eq. (11) can be re-written by expanding the weights and rearranging terms, resulting in

$$w(\mathbf{x}) \propto \frac{\exp -\beta U(\mathbf{x})}{\exp -\beta U_1(\mathbf{x})} . \qquad (12)$$

Eq. (12) shows that the total weight takes into account all the interactions missing in the non-interacting fragments described by the first intermediate $U_1$.

Note that the weights in Eqs. (11) and (12) are proportional to the ratio of the Boltzmann factors up to the constant which is the ratio of the corresponding partition functions. However, this constant is not needed for re-weighting because only the relative weights are important.

**II.D. Resampling**

In general, configurations with low weights have low importance in the ensemble and therefore it is desirable to save computer time by eliminating such configurations from future consideration. However, such elimination must be performed statistically to preserve the correct distribution [14]. Such a "resampling" process refers to eliminating, duplicating, and/or adjusting weights of configurations in the original ensemble resulting into an alternative ensemble [14]. Both ensembles are formally equivalent in representing the desired distribution.

A number of resampling algorithms have been suggested in statistics and data processing [14,44]. We implemented several resampling schemes in our growth algorithm and found a scheme termed "optimal resampling" [44] to be the most efficient. The advantage of optimal resampling compared to other schemes is that it guarantees distinct configurations and at the same time allows a large diversity of weights.

The main feature of optimal resampling is that it guarantees drawing the desired number of distinct configurations, denoted by $M$, from an original ensemble containing $N$ configurations and corresponding weights. This is achieved by employing a threshold weight $c$ which uniquely defines $M$. The configurations are accepted with probability $\min\left\{1, \frac{w_j(\mathbf{x})}{c}\right\}$, where $w_j(\mathbf{x})$ are the partial weights at stage $j$. The resampling of only distinct configurations is guaranteed by employing a special numerical cumulative distribution function (cdf) [44].

We implemented the optimal resampling in our growth algorithm at the end of each combination stage. After the fragments are joined and the weights are calculated, the configurations are resampled into a smaller ensemble containing 10% of the original configurations. The 10-fold reduction factor was found to be the most efficient based on trials of different $N$ and $M$ values. The typical ensemble size employed in our simulations is $N=10^5$ configurations, which is resampled into an ensemble of size $M=10^4$. As we describe in Sec. III.B, an "enrichment" procedure is employed to compensate for configurations eliminated by resampling and to maintain a constant ensemble size at different combination stages.

It is worth noting that after the last combination stage, configurations with weights may be resampled into an ensemble without weights. We implemented several different resampling algorithms to eliminate weights in the final ensemble. However, we consistently found that such resampling considerably reduces information contained in the weights. Therefore, after the last combination stage we use the same optimal resampling scheme as at other stages and save configurations with weights for further analysis. This is similar to keeping a larger number of correlated "snapshots" from a dynamics trajectory [45].

## II.E. Approximate ensembles

Besides equilibrium sampling, our growth procedure can be adapted for rapid generation of approximate ensembles. This may be useful for larger systems for which precise ensembles are not required – for instance, in schemes which assemble protein configurations from multi-residue segments [35,46-48]. The only new feature of our approximate procedure is that after the last combination stage configurations are used without weights. This way weights are used only to identify configurations without steric clashes. In other words, resampling works as a "bump check".

## II.F. Assessment of sampling precision and efficiency

In the present work efficiency of the growth algorithm is defined as the savings in wallclock time to achieve the same level of statistical precision in sampling of configuration space distribution relative to standard Langevin dynamics. This precision can be quantified by the number of statistically independent configurations contained in the trajectory (i.e., effective sample size (*ESS*)). To assess efficiency, time to generate a single statistically independent configuration can be compared between two methods. Thus, we define efficiency as

$$\gamma = \frac{t_{Langevin}}{t_{Growth}} \frac{ESS_{Growth}}{ESS_{Langevin}} \qquad (13)$$

where $ESS_{Growth}$ and $ESS_{Langevin}$ are the effective sample sizes of the growth and Langevin simulations respectively. The symbols $t_{Growth}$ and $t_{Langevin}$ denote wallclock times of growth and Langevin simulations respectively.

To calculate the *ESS* for both growth and Langevin simulations we used a recently developed statistical analysis [39]. Qualitatively, the idea is to divide configuration space into approximate physical states and calculate variance in each state. The variance is inversely proportional to the effective sample size. The approximate physical states can be constructed using Voronoi bins in configuration space [38]. The reference structures for the Voronoi procedure [49] are derived from the protocol described in Ref. [22].

To check the results of the previous method we also used a second method to calculate the *ESS* for Langevin simulations. This method employs our previously developed "de-correlation" time analysis and can be used only for dynamic simulations [22]. Briefly, the idea is to determine how much simulation time must elapse between configurations in the trajectory in order for them to exhibit the statistics of fully independent samples. Using the de-correlation time and the total simulation length the number of statistically independent configurations in the trajectory can be calculated.

**III. IMPLEMENTATION**

The growth formalism described in Sec. II does not lead to a unique algorithm, but can be implemented in many different ways. Implementation details are particularly important because modern forcefields are much more complicated than the early simple polymer models. Indeed, in our study we found that the efficiency of the growth algorithms depends significantly on the implementation. Here, we describe the technical approaches that helped to significantly improve the efficiency of our growth algorithm.

**III.A. Fragment libraries**

The advantage of using libraries is that some interactions and, therefore correlations within a molecule, can be calculated in advance and then used in multiple simulations saving CPU time. Instead of generating new fragment configurations on the fly, they can be cheaply retrieved from the memory. This approach is well suited for proteins which consist of only 20 different building blocks. We can build up libraries for different amino acids and then combine them according to the sequence to sample any peptide or protein. The idea to use molecular fragments in molecular simulations is well established in the literature [50,51] and has been successfully implemented in the protein structure prediction software Rosetta [35]. Earlier we have used libraries in a Monte Carlo approach [40].

Fragment libraries can be generated using any canonical method such as Langevin dynamics or Metropolis Monte Carlo. The only requirement for the libraries is that they should represent the true equilibrium distributions. In practice we used internal coordinate MC because it allows fixing some degrees of freedom such as some bond angles and dihedrals introduced with the dummy atoms. The dummy atoms were employed for two reasons. First they provide the six degree of freedom that specify the orientation of fragments relative to each. Second, the dummy atoms were chosen to interact with the real fragment atoms to provide better overlap with the full molecule distributions. We used libraries containing $10^5$ configurations.

We note that our fragments contain the same degrees of freedom and are sampled according to the same forcefield as employed in our previous study [38]. The only difference is that in our previous work the fragment libraries were generated by sampling the internal coordinates independently with subsequent reweighting into the full fragment distributions.

**III.B. Enrichment**

Enrichment entails making multiple copies of configurations at different stages of growth without introducing statistical bias, in order to increase the chances of partially grown chains to survive [6]. We implemented enrichment in our growth algorithm and found that it significantly increased the efficiency. One drawback of enrichment is that when chains are replicated, they are not longer statistically independent, limiting how much enrichment can ameliorate attrition. If chains are replicated too much, the configurations become too statistically correlated, and ultimately limit efficiency. We found that the most efficient implementation of enrichment in our growth algorithm is when it is applied after each combination stage and chains are replicated 10-100 times.

**III.C. Recycling of energy terms.**

In addition to coordinates, the potential energy of each fragment configuration can be calculated in advance and stored in the libraries. When fragments are combined, the potential energy of each fragment configuration can be cheaply retrieved from the computer memory saving CPU time. However, these savings will only be moderate for long molecules containing many fragments because interactions between fragments will dominate. We implemented recycling of energy terms in our growth algorithm and found that it helped to increase the efficiency for all the systems studied.

**III.D. Cartesian vs. internal coordinates**

To implement the growth formalism of Sec. II., it could seem natural to use internal coordinates, particularly for connecting fragments. However, each configuration ultimately must be converted to Cartesian coordinates for potential energy evaluation. In our original implementation fragment configurations were combined in internal coordinates and then converted to Cartesian for energy calculation. But we found that a large fraction of CPU time was actually spent on coordinate conversion.

The efficiency of our growth procedure was significantly improved when fragments were combined in Cartesian coordinates. This was implemented by storing "connector coordinates" – i.e. the six relative degrees of freedom – along with transformation matrices for each fragment configuration. First, the six degrees of freedom that specify the orientation of fragments relative to each other were used to set up the local coordinate systems. Given the local coordinate systems for each fragment, the appropriate transformation matrices were applied to generate the full Cartesian coordinates. In practice, configurations in the libraries were pre-oriented in the local coordinate system at the N-terminus of our residue based fragments and only one transformation matrix (at the C-terminus) was saved for each configuration in the library.

All transformation matrices were calculated using quaternion operations which allow fast and accurate transformations [52].

**III.E. Software optimizations**

The cost analysis of our growth algorithm revealed that it is "memory bound" – i.e., the bottleneck is not the CPU operations but rather the transfer of data from memory to CPU [53]. It is memory bound because it heavily relies on pre-calculating and storing configurations and energies in the memory. The transfer rate of data between the main memory and CPU is limited and becomes the bottleneck. To hide the memory latency problem modern CPUs utilize "cache" memory which allows much faster communication with CPU. However, the size of cache is much smaller than the main memory size so the data can be cached only in relatively small chunks. The memory bound algorithms can be improved by reusing the data and "neighbor use" [53]. Reuse helps to reduce the transfer of data from main memory to CPU by reusing as much as possible the data stored in cache and CPU registers. Neighbor use helps to perform computation on data (physically) close in memory reducing the transfer of data from memory to cache.

We implemented several standard optimization techniques in our C code including array linearization and blocking [53,54] both aimed at improving the reuse and neighbor use of fragment configurations and energies stored in the libraries.

**III.F. Breadth vs. depth**

The growth algorithm can be implemented in two different ways: "breadth first" and "depth first". In breadth first a whole ensemble of configurations is obtained at each intermediate stage before proceeding to the next one. In depth first only one full configuration is grown at a time. Both implementations have their own advantages and can be better suited for a particular resampling scheme etc.

Our implementation of the growth algorithm is a hybrid between breadth and depth. It is a hybrid because we grow a whole ensemble at once (typically $10^5$ configurations). However, to achieve a good statistical precision we repeat the whole growth process many times and simply combine configurations, energies and weights from different simulations into one large ensemble. Specifically, we used 10 repeats for Ace-$(Ala)_4$-Nme, 100 for Ace-$(Ala)_6$-Nme, and 1000 for Ace-$(Ala)_8$-Nme and Met-enkephalin. The approximate ensembles for Ace-$(Ala)_{12}$-Nme and Ace-$(Ala)_{16}$-Nme were generated using one repeat.

**IV. RESULTS**

We applied our polymer-growth algorithm to equilibrium sampling of several peptides including Ace-(Ala)$_4$-Nme, and Ace-(Ala)$_6$-Nme, Ace-(Ala)$_8$-Nme and Met-enkephalin. The equilibrium ensembles were sampled according to OPLS-AA forcefield [42] and for this initial study we used a simple solvent model with uniform dielectric of 60 at 298 K. The dielectric constant was chosen based on several trial simulations to give reasonable agreement for Ramachandran propensities with GBSA simulations [43]. As discussed in Sec. III.F. Ace-(Ala)$_4$-Nme was run for 10 repeated simulations resulting in $10^5$ saved structures, Ace-(Ala)$_6$-Nme was run for 100 repeats leading to $10^6$ configurations. Ace-(Ala)$_8$-Nme and Met-enkephalin were run for 1000 repeats also resulting in $10^6$ saved configurations.

To compare the growth results we ran standard Langevin dynamics simulations for the same four peptides described by the same forcefield and solvent model. Specifically, all systems were sampled for 200 ns at the temperature of 298 K and the friction constant of 5 ps$^{-1}$. The Langevin dynamics was used as implemented in Tinker software package [42]. All growth and Langevin dynamics simulations were performed on a single Xeon 3.6 GHz CPU and 2 GB of system memory.

We first checked that our algorithm can correctly sample the equilibrium distributions by comparing it with Langevin dynamics. The equilibrium distributions were compared using structural histograms constructed using Voronoi procedure as described in previous work [38]. The results are shown in

Figure 1 and indicate mostly good agreement between the two methods – although there appears to be slight bias in the Met-enkephalin results: see Discussion section.

To assess the efficiency of growth simulations we calculated the effective sample size (*ESS*) of Langevin simulations using two different statistical tools described in Sec. II.F. The first method is based on calculating the variance in the approximate physical states [39]. The second method employs our previously developed de-correlation time analysis [22] and was used to check the results of the first method which we recently developed [39]. The results are reported in Table 1 and indicate a reasonable agreement between two statistical methods. We note that the de-correlation time analysis can be used only for dynamic trajectories and, therefore, was not used for growth simulations.

The *ESS* of growth simulations was calculated using the first statistical tool i.e., by computing the variance in the populations of approximate physical states. The results of this analysis are reported in Table 2 and indicate that a large efficiency gain of over one order of magnitude was achieved for most peptides.

We emphasize that the efficiency of polymer growth algorithms applied to atomistic forcefields at 298 K depends significantly on implementation. In fact our original, naive implementation was not efficient at all – it was several times slower than Langevin simulations. However, in a series of implementation improvements described in Sec. III, we achieved good efficiency.

To aid future research in the field, we report how different improvements contributed to the efficiency of growing the peptide Ace-(Ala)$_4$-Nme. The largest improvement, of about two orders of magnitude, can be attributed to using Cartesian coordinates and recycling energy terms. Software optimizations improved the efficiency by about three times. Implementation of the optimal resampling algorithm increased the efficiency by almost another order of magnitude.

Besides equilibrium sampling of small peptides, our growth procedure can be also used for rapid generation of approximate ensembles (i.e., steric-clash free) of larger peptides. As described in Sec. II.E, we generated approximate ensembles for Ace-(Ala)$_{12}$-Nme and Ace-(Ala)$_{16}$-Nme peptides. Each required less than a minute of a single-processor wallclock time. To check whether these approximate ensembles provide reasonable distributions in configuration space, we compared them with equilibrium ensembles from Langevin simulations. The distributions were compared based on structural histograms constructed using a Voronoi procedure [38]. A larger number of bins (20) were used to investigate whether reasonable coverage of configuration space was obtained; the issue of coverage could be important in fragment-assembly applications. The results for both peptides are shown in Figure 2 and indicate that, indeed, the approximate ensembles yield reasonable coverage of configuration space.

## V. DISCUSSION

### V.A. Limitations

The key limitation of the present implementation of the growth algorithm is that it can be applied for precise equilibrium sampling only of relatively small peptides. The limit is about eight amino acid residues and less than 100 atoms, although we showed that significantly larger peptides can be treated approximately. The size limitation for precise sampling appears to result from the small overlap between non-interacting and fully interacting fragment distributions. In the present implementation, the overlap significantly decreases with system size because configurations which predominate in the partially grown ensemble may be less important in the full molecule. For example, if one is growing a largely helical peptide, partially grown configurations will not "know" about hydrogen bonds which will be formed later in the growth process. In Sec. V.C we describe possible solutions to the problem of small overlap.

The present procedure is also limited to implicit solvent models. It is not clear whether explicit solvent could be treated practically, but as discussed below, it should be possible in principle to grow explicit solvent.

**V.B. Possible applications**

There are numerous applications for any fast peptide sampling scheme, and some that are specific to the growth scheme. First, it is important to recognize that relatively small peptides have important functions as hormones, neurotransmitters, antigens, and drugs [55]. Examples include enkephalins and vasopressin, just to name a few. Pharmacologically, peptides are actively being investigated as potential drugs [56]. Fast growth procedures can permit the investigation of large numbers of sequences.

Unlike dynamic methods, the polymer growth approach can be used to calculate the absolute free energy without any additional cost [38]. This is possible because all the required generating probabilities and weights are known [57]. In our previous study [38], we calculated absolute free energies for several peptides based on pre-calculated molecular fragments; however, that study did not employ the critical efficiency improvements described here. In principle, different peptides (or other molecules) can be grown in the presence and absence of a protein to yield relative binding affinities via a standard thermodynamic cycle.

The approximate growth procedure could be of particular use in conjunction with fragment assembly protocols for protein structure prediction [46-48]. Presently, these protocols rely on expensive dynamic sampling of fragment configurations for subsequent assembly into native-like structures. Our growth procedure can rapidly generate approximate ensembles of peptides suitable for such assembly or perhaps with other fragment-based programs like Rosetta [35,58].

Interestingly, the growth procedure can be adapted to multi-resolution sampling because of flexibility in how a molecule is divided into fragments. For example, fragments can correspond to the backbone and side chains of different types. In such a version of the growth algorithm – that we will call "decorating" [31] – given a backbone ensemble, side chains can be added one at a time. Decorating is a true multi-resolution technique because the backbone can be sampled using other canonical methods, for example, our previously developed library-based Monte Carlo [40]. Initial data obtained from decorating (not shown) suggest it can be a useful approach.

**V.C. Possible improvements**

There are several possible solutions to the problem of small overlap. One possibility is to bias the growth based on some prior knowledge of the full molecule's configuration-space distribution. This knowledge may be obtained from previous dynamics or growth simulations even if these simulations are not fully sampled, provided there is some information on correlations among all fragments. For example, the biasing can be implemented as a "self-guided" algorithm: a regular growth simulation can be performed first and then subsequent growth simulations can be biased toward important parts of configuration space based on the information obtained in the first simulation. Libraries could also be biased based on simulations and/or databases like the protein data bank.

Efficiency for larger systems might be improved by expanding the ensemble at every intermediate stage $j$. For instance, ensemble expansion could be effected using local "relaxation" of the growing configurations with a canonical sampling method, such as library-based Monte Carlo [40]. This idea is based on "annealed importance sampling" [59,60]. An enlarged canonical ensemble at stage $j$ should have more configurations pertinent to stage $j$+1. In general, growth and dynamic approaches have features that can help each other to better sample configuration space. Growth can instantaneously cross potential energy barriers but is not good at exploring local configuration space. On the other hand, relaxation of partially grown configurations may help to remove strains and better explore local configuration space. Canonical relaxation preserves the correct distribution [59-61].

It is natural to consider fragments larger than those used here, which were single amino acids. For instance, fragments can correspond to amino-acid dimers or even larger peptides. There are two practical limitations on fragment size, however, both of which will become less severe with increasing memory. One restriction stems from sequence variations within a fragment. For example, for dimer fragments it will be necessary to generate and store at least $\frac{20 \times 19}{2} = 190$ different libraries. Another practical limitation is the number of configurations required to adequately describe each library. Our present procedure employs $10^5$ configurations per library but larger fragments may require significantly larger libraries. On the other hand, biasing libraries toward the most pertinent parts of configuration space will decrease the storage requirements. Again, as computer memory increases and becomes more affordable, these limitations may become less important.

Despite these limitations we tested the potential of using larger fragments in our growth procedure. We employed $(Ala)_2$ and $(Ala)_3$ fragment libraries each containing $10^5$ configurations to sample Ace-$(Ala)_8$-Nme but found that the efficiency was inferior compared to using a single Ala fragment. One reason why larger fragments did not help may be that they require much larger libraries (compared to single-residue fragments) to represent their large configuration space.

In this initial study we employed a simple solvent model with uniform dielectric although more accurate models such as GBSA [43] can be implemented. When using a new solvent model, fragment libraries will have to be regenerated although it requires only one time cost. Additional energy terms for the solvent model will have to be implemented in the algorithm.

In principle, polymer growth algorithms are not limited to implicit solvent models. Similar to growing peptides, water molecules can be added one at a time to solvate the system. In fact, our group has already "grown" a simple Lennard-Jones fluid [62].

The polymer growth algorithms are well suited for modern graphics processing units (GPUs) because multiple configurations can be grown at once in contrast to dynamic simulations where only one configuration can be processed at a time. The advantage of GPUs is that they have hundreds of arithmetic units where multiple interactions and/or configurations can be simultaneously processed.

## VI. CONCLUSIONS

We report the use of a polymer-growth algorithm that employs pre-calculated molecular fragment libraries for equilibrium sampling of peptides using an atomistic forcefield (OPLS-AA) at 298 K. To authors' knowledge this is the first application of the polymer-growth technique for equilibrium sampling of atomistic protein models at a semi-physiological temperature. The results show that the approach is correct and can be considerably more efficient than standard Langevin dynamics for several implicitly solvated peptides. Approximate ensembles for larger peptides (up to Ace-(Ala)$_{16}$-Nme in the present study) can be generated in less than a minute of single-processor computing time.

The considerable speed of the calculations can be attributed to the implementation of several optimization techniques, some of which are not applicable to standard dynamics methods. Future improvements such as biased libraries, self-biasing, and relaxation may help to further improve speed and efficiency, especially for large systems. Our results seem to warrant further studies of the polymer growth strategy for equilibrium sampling of polypeptides.

## Acknowledgment

The authors would like to thank Ed Lyman, Ying Ding, and Divesh Bhatt for helpful discussions. Funding was provided by the NIH through Grants GM070987 and GM076569, as well as by the NSF through Grant MCB-0643456.

**Figures**

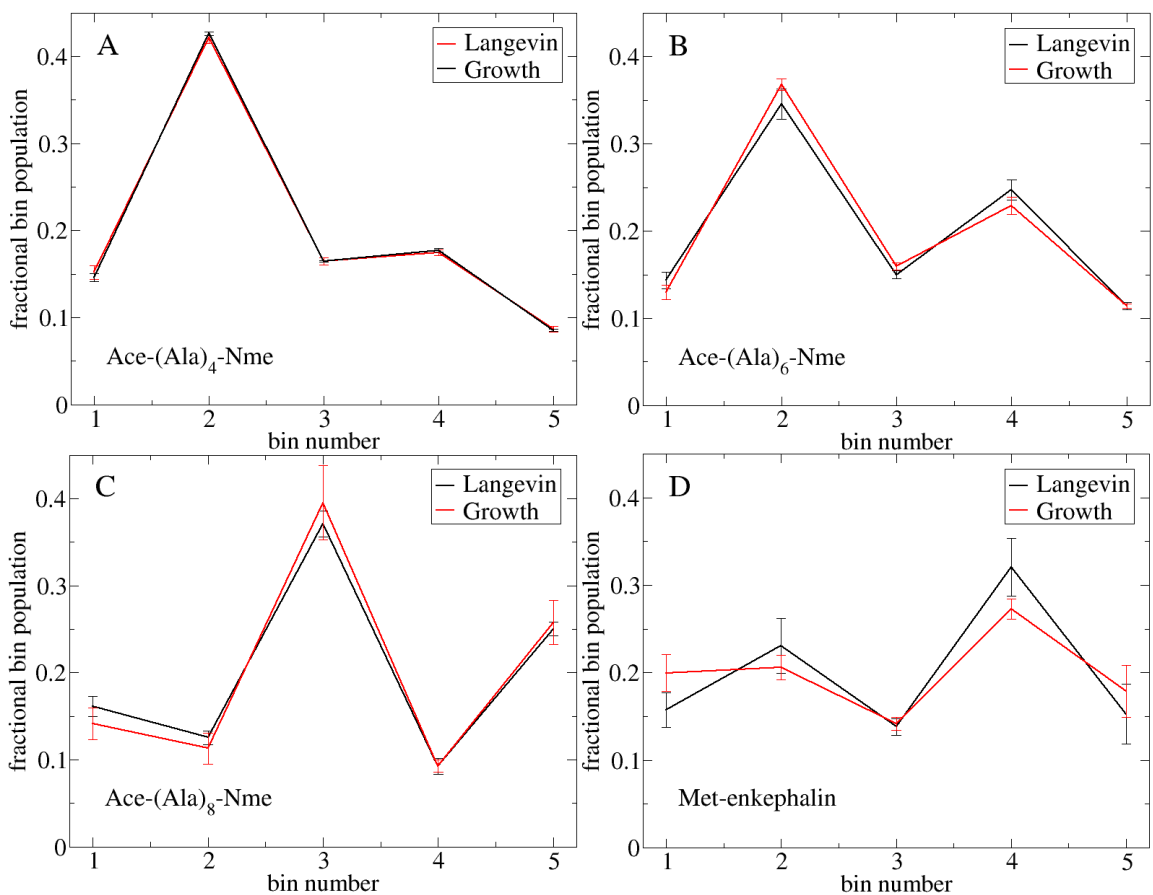

**Figure 1.** Fractional population of Voronoi bins constructed from growth and Langevin simulations for four peptides: (A) Ace-(Ala)$_4$-Nme, (B) Ace-(Ala)$_6$-Nme, (C) Ace-(Ala)$_8$-Nme, and (D) Met-enkephalin. The bins were constructed based on a Voronoi classification of configuration space as described in Ref. [38]. Error bars represent one standard deviation for each bin, based on 12 independent simulations for both Langevin and growth.

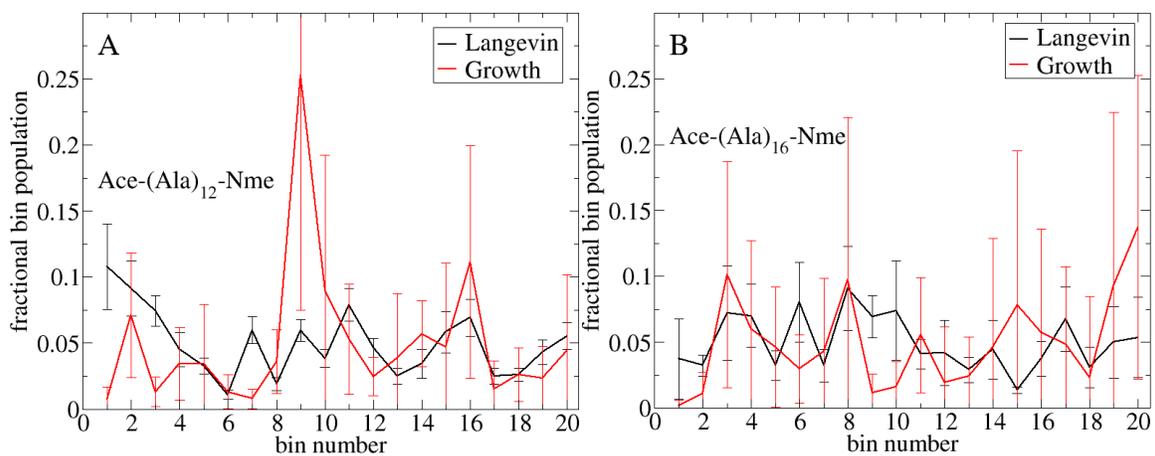

**Figure 2.** Fractional populations of Voronoi bins constructed from approximate growth procedure and Langevin simulations for two peptides: (A) Ace-(Ala)$_{12}$-Nme, and (B) Ace-(Ala)$_{16}$-Nme. The bins were constructed based on a Voronoi classification of configuration space as described in Ref. [38]. Error bars represent one standard deviation for each bin, based on 12 independent simulations for growth and 10 for Langevin.

# Tables

**Table 1.** The results of statistical analysis of Langevin dynamics simulations are reported for four peptides. The effective sample size ($ESS_{Langevin}$) was calculated using two different statistical tools as described in Sec. II.F.

| System | Number of Atoms | $t_{Langevin}$ | $ESS_{Langevin}$ from physical states analysis | $ESS_{Langevin}$ from de-correlation analysis |
|---|---|---|---|---|
| Ace-(Ala)$_4$-Nme | 52 | 28 h | 2180 | 2500 |
| Ace-(Ala)$_6$-Nme | 72 | 48.3 h | 615 | 800 |
| Ace-(Ala)$_8$-Nme | 92 | 76 h | 385 | 330 |
| Met-enkephalin | 84 | 80 h | 55 | 130 |

**Table 2.** The results of the statistical analysis of growth simulations are reported for four peptides. The effective sample size ($ESS_{Growth}$) was obtained based on calculating the variance in the approximate physical states as described in Sec. II.F. The efficiency gain $\gamma$ relative to Langevin dynamics was calculated using Eq. (13). Note that $\gamma$ was obtained using $ESS_{Langevin}$ calculated from the variance in the physical states.

| System | Number of Atoms | Number of Fragments | $t_{Growth}$ | $ESS_{Growth}$ | $\gamma$ |
|---|---|---|---|---|---|
| Ace-(Ala)$_4$-Nme | 52 | 6 | 1 min | 2800 | 2150 |
| Ace-(Ala)$_6$-Nme | 72 | 8 | 10.6 min | 170 | 75 |
| Ace-(Ala)$_8$-Nme | 92 | 10 | 1.75 h | 45 | 5 |
| Met-enkephalin | 84 | 7 | 1.5 h | 100 | 100 |